\documentclass{elsart}
\usepackage{natbib}
\usepackage{epsfig}
\input{epsf.sty}
\usepackage{epsfig}
\usepackage{longtable}

\begin{document}
\runauthor{Xu}
\begin{frontmatter}

\title{To probe into pulsar's interior through gravitational waves}

\author[Beijing]{R. X. Xu}\footnote{
Corresponding author.\\
{\em Email address:} r.x.xu@pku.edu.cn.}

\address[Beijing]{School of Physics, Peking University,
Beijing 100871, China}

\begin{abstract}
The gravitational radiation from compact pulsar-like stars depends
on the state of dense matter at supranuclear densities, i.e., the
nature of pulsar (e.g., either normal neutron stars or quark
stars).
The solid quark star model is focused for the nature of
pulsar-like compact objects.
Possible gravitational emission of quark stars (either fluid or
solid) during the birth and later lifetime is discussed. Several
observational features to distinguish various models for
pulsar-like stars are proposed.
It is suggested that the gravitational wave behaviors should be
mass-dependent. Based on the data from the second LIGO science
run, the upper limits of $R\cdot \theta^{1/5}$ and thus $M\cdot
\theta^{3/5}$ ($M$: mass, $R$: radius, and $\theta$: wobble angle)
are provided for those targets of millisecond pulsars.

\vspace{5mm} \noindent {\it PACS codes:} 97.60.G, 97.60.J,
11.80.F, 95.85.S
\end{abstract}


\end{frontmatter}


\section{Introduction}

On the one hand, it is now a very critical time for detecting {\em
directly} gravitational waves.
Despite the negative detection results, some limits on the
physical parameters of pulsar-like star's interiors can be
obtained with theoretical calculations which relate the waves to
star's nature in literatures, based on the experiment data of
gravitational waves.
Andersson \& Kokkotas \citep{ak98} calculated the eigenfrequencies
of the $f$, $p$, and $w$ modes, and found that the frequencies
depend on star's mass and radius.
Numerical results \citep{bbf99} showed that the information of
neutron star structure could be carried by gravitational waves of
the axial $w$-modes.
The inverse problem for pulsating neutron stars, based on the
studies of those modes, was discussed extensively by Kokkotas,
Apostolatos \& Andersson \citep{kaa01}.
The frequencies and damping behaviors were reexamined for
different equations of state \citep{bfg04}, and recently, Tsui \&
Leung \citep{tl05} applied an inversion scheme to determine star's
mass, radius, and density distribution via gravitational wave
asteroseismology.
From the theoretical work above, one sees then that a positive
result of recording gravitational wave signal should improve
significantly the knowledge of matter at supranuclear
density\footnote{
It may be worth noting, however, that gravitational wave or
graviton might not exist at all in the de Sitter Universe
\citep{liu04} if gravitational wave would still not be detected in
the next decades.}.

On the other hand, pulsar-like compact stars were discovered since
1968, but, unfortunately, we are still not sure about their
nature. It is conventionally believed that these compact stars are
normal neutron stars. However one can {\em not} rule out the
possibility, that they are actually quark stars composed of quark
matter with de-confined quarks and gluons, either from the first
principles or according to their various observations
\citep{glen00,lp04,weber05}.
Quarks are fundamental Fermions in the standard model of particle
physics, and the existence of quark matter is a direct consequence
of the asymptotically free nature of the strong interaction, which
was proved in the regime of non-Abelian gauge theories and
confirmed by high-energy collider experiments.
In this sense, to affirm or negate the existence of quark stars is
of significantly fundamental meaning for understanding the
nature's elementary strong interaction.
Quark stars with strangeness are popularly discussed in
literatures, which are called as strange (quark) stars, whereas
nonstrange quark stars would also be possible if nonstrange quark
matter could also be stable at zero pressure due to color
confinement \citep{mm05}. We do not differentiate between the
terms of quark stars and strange stars here, supposing both kinds
of the stars have a quark surface to confine quarks and gluons.

Therefore, one of the key points for today's astrophysicists is
understanding the nature of pulsars: to find competitive evidence
for quark stars or for normal neutron stars.
Effective methods to do depend on (i) the minimum spin periods,
(ii) the mass-radius relations, and (iii) the surface differences
\citep{xu03b}.
However, since they are relativistic, pulsar-like compact stars
could be potentially gravitational wave radiators. The features
and strength of this emission would reflect their nature, and one
may probe into pulsar's interior with gravitational waves.
Recently, the Laser Interferometric Gravitational wave Observatory
(LIGO) has monitored 28 radio pulsars \citep{LIGO2005}, and
obtained upper limits of gravitational radiative intensities for
the pulsars.
If pulsars are quark stars, how and what can one constraint the
pulsar's nature based on the data? We are trying to analyze
relevant issues in this paper.

There are actually two motivations to study gravitational waves
using pulsars: (i) as a tool and (ii) as a source.
These two result in different physics. The radio pulsar timing
array depends on the ``clock'' nature, not on the internal
structure of pulsars, while the direct detection of the waves from
isolated and/or binary pulsars depends on the internal structure
of pulsars. The former is not discussed in this paper.
Certainly, a successful detection of gravitational waves from
pulsar-like stars is very important for understanding (i) the
physics of matter at supranuclear density and (ii) the physics of
gravity.

\section{Quark stars during their birth}

A mass formula for strangelet analogous to the Bethe-Weizsacher
semi-empirical mass function in nuclear physics was introduced by
\cite{dcs93}. Neglecting the symmetry and Coulomb energy terms,
which are not important in the case of strange quark stars, one
has $E=E_{\rm v}+E_{\rm s}$, where the volume and surface energy
contributions are $E_{\rm v}=\epsilon_0A'$ and $E_{\rm s}\sim
0.1\epsilon_0A'^{2/3}$, respectively, with $\epsilon_0=(880\sim
890)$ MeV and $A'$ the baryon number \citep{dcs93}.
By introducing baryon number density $n_{\rm b}=A'/(4\pi R^3/3)$
($R$ the radius of spherical quark matter) and the surface energy
per unit area $\gamma$, we have $R=(3A'/(4\pi n_{\rm b}))^{1/3}$
and thus
$\gamma\simeq 0.02\epsilon_0n_{\rm b}^{2/3}\simeq 20$
MeV/fm$^2$.
However, \cite{bagchi05} obtained a surface energy (or surface
tension), $\gamma$, which varies from about 10 to 140 MeV
fm$^{-2}$, depending on stellar radius.
It is worth noting that the surface energy, which will be
neglected in this paper, due to an electrostatic field of $<\sim
10^{17}$ V/cm on a star's surface is several orders smaller that
10 MeV fm$^{-3}$.
We consider thus this poorly known parameter $\gamma$ to range
between $\sim 10^1$ MeV fm$^{-2}$ and $\sim 10^2$ MeV fm$^{-2}$.

\subsection{Quark nuggets ejected}

\begin{figure}
\centerline{\psfig{file=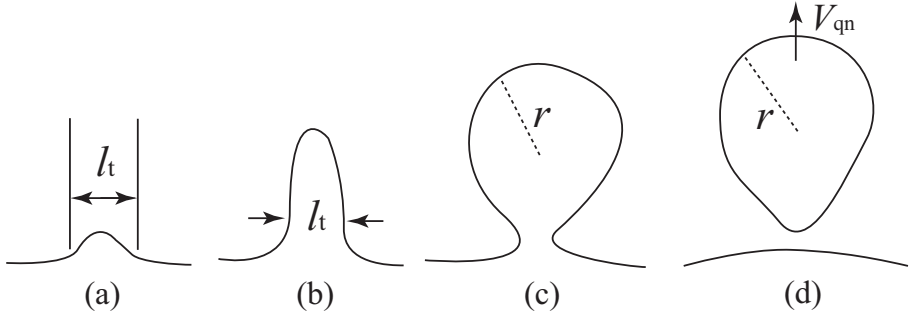,width=12cm}}
\caption{Forth steps to eject a quark nugget during the formation
of a quark star.
\label{birth}}
\end{figure}

A hot turbulent bare quark star may eject (or ``evaporate'')
low-mass quark matter (quark nuggets). Phenomenologically, there
could be 4 steps to create a quark nugget (Fig.\ref{birth}).
Two parameters, which are not know with certainty yet, describe
the turbulent nature of protostrange stars: the convective scale
$l_{\rm t}$ and velocity $v_{\rm t}$. It is possible that $l_{\rm
t}\sim 10^{2\sim 5}$ cm, $v_{\rm t}\sim 10^5$ cm/s \citep{xb01}.
In step (a), the typical kinematic energy of turbulent part is
$E_{\rm t}\sim \rho l_{\rm t}^3 v_{\rm t}^2$, while the surface
and the gravitational energies, which may prevent the fluid part
to eject, are $E_{\rm s}\sim \gamma l_{\rm t}^2$ and $E_{\rm
g}\sim GmMr/R^2$ (quark star's mass $M$ and its radius $R$, lump's
mass $m$ and radius $r$), respectively.
The height of a bulge increases if $E_{\rm s}+E_{\rm g}<E_{\rm
t}$, as is shown in step (b).
The contact area between the star and the bump becomes smaller and
smaller, as shown in step (c), because of (i) a limited scale of
kinematic flow with order, and (ii) a total momentum which tends
to separate nugget from star. A quark nugget form finally, with a
velocity of $V_{\rm qn}<\sim v_{\rm t}$, step (d).
The density of a bare strange star with mass $\leq \sim M_\odot$
is nearly uniform \citep{afo86}, and the mass of such a star can
be well approximated by,
\begin{equation}
M\simeq {4\over 3}\pi R^3(4B),%
\label{mr}
\end{equation}
where the bag constant $B=(60\sim 110)$ MeV/fm$^3$, i.e.,
$(1.07\sim 1.96)\times 10^{14}$ g/cm$^3$.
For simplicity, we apply this approximation throughout this paper.

What kind of quark nugget could be ejected? Let's compare the
energies below,
\begin{equation}
{E_{\rm t}\over E_{\rm g}}\sim {R^2v_{\rm t}^2\over GMr}\simeq
8\times 10^{-15}R_6^{-1}{v_{\rm t}^2\over r},~~{E_{\rm t}\over
E_{\rm s}}\sim {\rho_0v_{\rm t}^2r\over \gamma}\simeq 2\times
10^{-8}\gamma_{100}^{-1}v_{\rm t}^2r,
\label{compare}
\end{equation}
where $\rho_0\simeq 4\times 10^4$ g/cm$^3$ is the average density
of quark stars with mass $< \sim M_\odot$, $R=R_6\times 10^6$ cm,
$\gamma=\gamma_{100}\times 100$ MeV/fm$^2$. Both $E_{\rm t}/E_{\rm
g}$ and $E_{\rm t}/E_{\rm s}$ should be much larger than 1 if
quark nuggets are ejected.
We conclude then from Eq.(\ref{compare}) that: (i) it needs
$v_{\rm t}>10^8$ cm/s that nuggets with mass $>10^{20}$ g can be
ejected; (ii) nuggets with higher mass could be ejected from
nascent quark stars with lower mass; (iii) ejection of nuggets
with planet masses (e.g., the Earth's mass $\sim 10^{27}$ g) could
be possible if $v_{\rm t}>10^9$ cm/s; (iv) a nascent quark star
with weak turbulent (e.g., $v_{\rm t}<10^4$ cm/s) could hardly
evaporate nuggets with low mass (e.g., $<10^{14}$ g).
Hadrons may not evaporate from the surface of a quark star with
temperature being much lower than 100 MeV if $\gamma$ is order of
100 MeV/fm$^2$.
The unknown parameter $v_{\rm t}$ would be probably much larger
than $10^7$ cm/s since observations show that pulsar-like stars
receive a large kick velocity (of order a few hundred to a
thousand km/s) at birth.

These ejecta could be captured by the center strange star (with
mass $M$ and radius $R$) for $v_{\rm t}$ should be smaller than
the escape velocity $\sqrt{2GM/R}\sim 10^{10}$ cm/s.
We then conclude that {\em quark planets} (i.e., planets of quark
matter, with possible mass from much lower than earth mass to
about a Jupiter mass) would form simultaneously during the
formation of a strange star with strong turbulence if the surface
energy is reasonable ($\gamma < 10^3$ MeV fm$^{-3}$).
This consequence could be tested by searching quark planets around
pulsars.
This is also an alternative mechanism for creating pulsar-planet
systems, the first one of which was discovered by \cite{wolszc92}.

\subsection{Global oscillation of quark stars}

A spherical fluid should be deform by rotation. For a quark star
of incompressible fluid with axis-symmetric deformation, the
star's surface can be described as,
\begin{equation}
r(\theta)=\lambda R[1-\sum_{i=2}^n\alpha_iP_i(\cos\theta)],
\label{r}
\end{equation}
where the parameter $\lambda$ is determined by a constant volume
of $4\pi R^3/3$. In case of $\alpha_2\ll 1$ (and thus terms with
orders being higher than $\alpha_2^2$ are neglected), the
quadrupole deformation ($i=2$), which will only be noted as
following, is actually of an ellipsoidal figure with an
ellipticity $\varepsilon=\alpha_2$ and axes
$a=R/\sqrt{1-\varepsilon}=b>c=R(1-\varepsilon)$.

In case of a rotating fluid star with $T/|W| \ll 0.1375$, the
ellipticity $\varepsilon\equiv(I-I_0)/I_0$ and the eccentricity
$e=1-c^2/a^2$ are related by $\varepsilon\approx e^2/3$, where
$I(\Omega)$, as a function of the spin frequency $\Omega$, is the
total moment of inertia, with $I_0=2MR^2/5$ for stars with uniform
density.
The total energy, $E$, of a rotating fluid star is then the sum of
gravitation, rotation, volume, and surface energies. A calculation
similar to the nuclear liquid drop model\footnote{%
The Coulomb interaction there is replaced by gravitational
interaction, with a sign-change.} \citep{gm96} shows
\begin{eqnarray}
E &=& E_{\rm gravi} + E_{\rm v} + E_{\rm s} + E_ {\rm
rot}\nonumber\\ &=& E_{\rm 0} + (A_g + A_s) \varepsilon^2 +
{L^2\over 2I_0}{1\over (1+\varepsilon)},
\label{E}
\end{eqnarray}
where $E_{\rm 0}$ is the energy for a non-rotating spherical star,
$L=I\Omega$ the star's angular momentum, $I(\varepsilon)$ the
momentum of inertia, and the coefficients $A_g$ and $A_s$ measure
the increases of gravitation and surface energies, respectively,
of the star,
\begin{equation}
A_g = \frac{3}{{25}}\frac{{GM_{}^2 }}{R},~~~A_s = {8\pi\over
5}\gamma R^2.
\end{equation}
The surface energy contribution to the deformation should be
negligible since the parameter
\begin{equation}
\eta\equiv {A_s\over A_g}\simeq {15\over 2\pi G\rho_0^2}\cdot
{\gamma\over R^3}\sim 10^{-18}\gamma_{100}R_6^{-3}
\label{eta}
\end{equation}
is much smaller than 1 unless the star is very low-massive (i.e.,
mass $<\sim 10^{14}$ g for strangelets) and/or $\gamma\gg 100$
MeV/fm$^2$.

By using ${\rm d}E(\varepsilon)/{\rm d}\varepsilon=0$ with
Eq.(\ref{E}), neglecting $A_s$, one comes to the rotating figures
of conventional Maclaurin spheroids\footnote{%
Note that the approximation of $\varepsilon_0(P)$ can only be
valid for $P_{\rm 1ms}\gg 1$.
},%
\begin{equation}
\Omega\simeq 2\sqrt{2\pi G\rho_0\over 15}e_0, ~~{\rm or,}%
~~\varepsilon_0(P)\simeq {5\Omega^2\over 8\pi G\rho_0}\simeq 3\times 10^{-3}P_{\rm 10ms}^{-2},%
\label{varepsilon0}
\end{equation}
where $e_0$ and $\varepsilon_0$ are the eccentricity and the
ellipticity, respectively, of a fluid star at equilibrium, which
are only a function of spin, provided that $\rho_0$ is a constant,
the spin period $P=2\pi/\Omega=P_{\rm 10ms} \times 10$ ms.
One can also have
\begin{equation}
{{\rm d}^2E(\varepsilon)\over {\rm d}\varepsilon^2}=
2(A_g+A_s)+{L^2\over I_0(1+\varepsilon)^3}>0,
\end{equation}
which means the equilibrium state of $\varepsilon=\varepsilon_0$
is stable.
We expect that a fluid quark star, either at birth or in its
lifetime, should {\em oscillate} globally around $\varepsilon_0$,
though the oscillation frequency and amplitude are still not
computed now. If the amplitude is large, this quadrupole
oscillation could result in enough gravitational radiation to be
detected by future advanced facilities.

Additionally, this oscillation modes may help us to distinguish
fluid and solid quark star models.
A protoquark star should be in a fluid state, but the star may
solidified soon after its birth \citep{xu03a,xu05b}. Therefore, we
should find ways to differentiate solid and fluid quark stars in
their later time. Due to the strong shear force, a solid quark
star may oscillate with much smaller amplitude but much higher
frequency than that of a fluid quark star. Besides detecting the
gravitational wave, pulsar timing in radio, optical, as well as
X-ray bands could be effective if the star has pulsed emission,
because the rotation period changes during the oscillation.

Could the oscillation amplitude be larger enough to fragmentate a
fluid quark star?
Unlike the case of nuclear fission, where the Coulomb interaction
favors, the gravitational interaction prevent a quark star to
fragmentate. It needs then energy to ``excite'' the star, with
order of $E_{\rm ex}\sim 0.6G(M/2)^2/R$. If pulsar kick
\citep{popov04}, with energy of $E_{\rm ki}\sim 0.5MV_k^2$, plays
a major role to fission a quark star, we note that the kick
velocity $V_k$ should be greater than $\sim R\sqrt{2\pi
G\rho_0/5}\simeq 6\times 10^3R$ cm/s.
Therefore, a quark star with large kick (or turbulent) energy and
low mass could fragmentate.

\section{Gravitational waves due to $r$-mode instability}

In Newtonian theory, a rapidly rotating fluid Maclaurin spheroid
is secularly unstable to become a Jacobi spheroid, which is
non-axisymmetric, if the ratio of the rotational kinetic energy to
the absolute value of the gravitational potential energy $T/|W|
> 0.1375$.
In the general relativistic case, as was shown by \cite{chandra70}
and \cite{fs78}, gravitational radiation reaction amplifies an
oscillation mode, and it is then found that the critical value of
$T/|W|$ for the onset of the instability could be much smaller
than 0.1375 for neutron stars with mass $>\sim M_\odot$.
This sort of non-axisymmetric stellar oscillations will inevitably
result in gravitational wave radiation.

A kind of oscillation mode, socalled $r$-mode, is focused on in
the literatures since the work by \cite{and98}, \cite{fm98}, and
\cite{lom98}.
The $r$-mode oscillation is also called as the Rossby waves that
are observed in the Earth's ocean and atmosphere, the restoring
force of which is the Coriolis force.
This instability may increase forever if no dissipation occurs.
Therefore, whether the instability can appear and how much the
oscillation amplitude is depend on the interior structure of
pulsars, which is a tremendously complicated issue in supranuclear
physics.

To make sense of the generic nature for $r$-mode instability in
different star-modes but avoiding an uncertainty in microphysics,
we just present a rough calculation below.
The critical angular frequency, $\Omega$, limited by the
gravitational radiation due to $r$-mode instability, are
determined by comparing the damping/growth times due to the
various mechanisms for energy dissipation in a given star-model,
\begin{equation}
{1\over \tau_{\rm gw}}+{1\over \tau_{\rm sv}}+{1\over \tau_{\rm
bv}}=0,
\label{r-mode}
\end{equation}
where the timescales for the instability are estimated to be
\citep{mad98},
\begin{equation}
\left\{
\begin{array}{lll}
\tau_{\rm gw} & = & -3.85\times 10^{81}\Omega^{-6}M^{-1}R^{-4},\\
\tau_{\rm sv}^{\rm ns} & = & 1.01\times 10^{-7}M^{-1}R^5T^2,\\
\tau_{\rm bv}^{\rm ns} & = & 1.29\times
10^{61}\Omega^2M^{-1}R^5T^{-6},\\
\tau_{\rm sv}^{\rm ss} & = & 1.85\times 10^{-9}\alpha_{\rm s}^{5/3}M^{-5/9}R^{11/3}T^{5/3},\\
\tau_{\rm bv}^{\rm ss} & = & 5.75\times 10^{-2}m_{100}^{-4}\Omega^2R^2T^{-2},%
\end{array}
\right. %
\label{taudiss}
\end{equation}
where superscript ``ns'' (``ss'') denotes neutron (bare strange)
star model, $\tau_{\rm sv}$ and $\tau_{\rm bv}$ are the
dissipation timescales due to shear and bulk viscosities,
respectively, and $\alpha_{\rm s}$ the coupling constant of strong
interaction, $T$ the temperature, $m_{100}$ the strange quark mass
in 100 MeV.
It is assumed in Eq.(\ref{taudiss}) that the modified URCA process
dominate the weak interaction in normal neutron stars.

Strange stars could have very low masses, even of $10^{-3}M_\odot$
(a strange star with planet mass could also be called as a strange
{\em quark} planet).
A low-mass normal neutron star could not be likely because of (i)
the minimum mass of a stable neutron star is $\sim 0.1M_\odot$
\citep{st83}, and
(ii) the gravitation-released energy of possible low-mass neutron
star during an iron-core collapse supernova, $E_{\rm g}\sim
GM^2/R$, should be much smaller than $\sim 10^{53}$ erg due to the
relation $M\sim R^{-3}$ for low-mass neutron stars, but it seems
not successful in modern supernova simulations even in case of
$E_{\rm g}\sim 10^{53}$ erg \citep{janka03,lieb04}.
Therefore, in the following calculations, we consider low mass
issues for bare strange stars, but apply only canonical mass of
$1.4M_\odot$ for normal neutron stars.

No $r$-model instability occurs in a solid star since the
Coriolis-restoration force does not work here.
The break frequency of low-mass bare strange stars could be
approximately a constant,
\begin{equation}
\Omega_0=\sqrt{GM\over R^3}=1.1\times 10^4~~{\rm s^{-1}},
\end{equation}
with a prefactor of $\sim 0.65$ at most for $M\sim M_\odot$ and
$R\sim 10^6$ cm \citep{glen00}, where the bag constant $B=60$
MeV/fm$^3$.
Let's express the critical frequency in unit of $\Omega_0$,
through calculations based on Eq.(\ref{r-mode}).
\begin{figure}
\centerline{\psfig{file=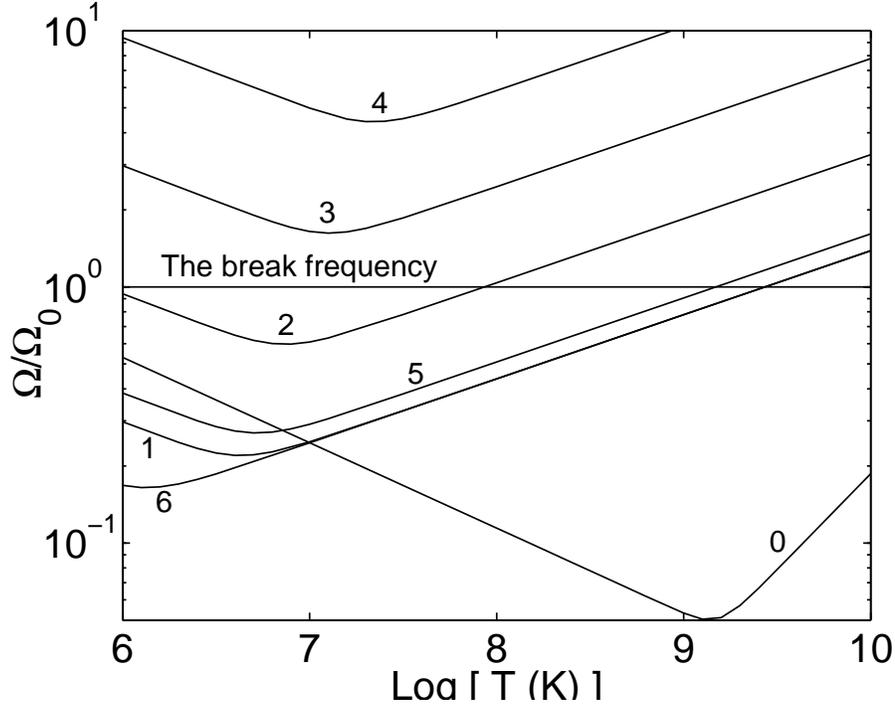,width=12cm}}
\caption{Temperature dependence of the critical angular frequency,
in unit of $\Omega_0$, due to $r$-mode instability for normal
neutron stars and bare strange stars.
The critical lines are numbered from ``0'' to ``6''. ``0'': normal
neutron stars; ``1'': bare strange stars (BSSs) with mass
$M=M_\odot$, the bag constant $B=60$ MeV/fm$^{-3}$, the strange
quark mass $m_{\rm s}=100$ MeV, and the coupling constant
$\alpha_{\rm s}=0.1$; ``2'': BSSs with $M=0.1M_\odot$, $B=60$
MeV/fm$^{-3}$, $m_{\rm s}=100$ MeV, and $\alpha_{\rm s}=0.01$;
``3'': BSSs with $M=0.1M_\odot$, $B=60$ MeV/fm$^{-3}$, $m_{\rm
s}=100$ MeV, and $\alpha_{\rm s}=0.1$; ``4'': BSSs with
$M=0.001M_\odot$, $B=60$ MeV/fm$^{-3}$, $m_{\rm s}=100$ MeV, and
$\alpha_{\rm s}=0.1$; ``5'': BSSs with $M=M_\odot$, $B=110$
MeV/fm$^{-3}$, $m_{\rm s}=100$ MeV, and $\alpha_{\rm s}=0.1$;
``6'': BSSs with $M=M_\odot$, $B=60$ MeV/fm$^{-3}$, $m_{\rm
s}=100$ MeV, and $\alpha_{\rm s}=0.9$.
}%
\label{rmode}
\end{figure}
It is found in Fig.\ref{rmode} that: (i) gravitational wave
radiates more likely from proto-neutron stars than from
proto-strange stars; (ii) the $r$-mode instability could not occur
in fluid bare strange stars with radii being smaller than $\sim 5$
km (or mass of a few $0.1M_\odot$) unless these stars rotates
faster than the break frequency; and (iii) the above conclusions
do not change significantly in the reasonable parameter-space of
$B$, $\alpha_{\rm s}$, and $m_{\rm s}$.
To estimate strength of gravitational waves from this oscillation,
one needs to simulate the nonlinear increase of the instability,
which is not certain yet.

Some recent observations in X-ray astronomy could hint the
existence of low-mass bare strange stars \citep{xu05a}.
The radiation radii (of, e.g., 1E 1207.4-5209 and RX J1856.5-3754)
are only a few kilometers.
No gravitational wave emission could be detected from such fluid
stars even they spin only with a period of $\sim 1$ ms.

\section{Gravitational waves from wobbling stars}

Gravitational wave radiation from pulsars could be classified as
(i) emission due to the normal modes of oscillation of fluid
matter (e.g., the $r-$mode instability discussed in \S3), and (ii)
emission due to solid deforming which will be focused in this
section.
%
%
Protoquark stars should be in a fluid state when their
temperatures are order of 10 MeV, but would be solidified as they
cool to very low temperatures \citep{xu03a}.
Assuming the ellipticity of a solid quark star, with an initial
spin period $P_0$, keeps the same as that of the star just in its
fluid phase, we expect realistic ellipticity $\epsilon$ of the
star in the solid state satisfies
\begin{equation}
\varepsilon_0(P_0)>\epsilon>\varepsilon_0(P)%
\label{epsilon}
\end{equation}
as the star spins down to a period of $P$, due to the shear force
that prevent the star to deform.
For the Earth which is spinning down, the mean density is 5.5
g/cm$^3$, the ellipticity observed $\epsilon=0.00223$, one has
$\varepsilon_0$(24 hours)$=0.00287>\sim \epsilon$ from
Eq.(\ref{varepsilon0}). Additionally, a free precession mode
called the Chandler wobble, which was formulated and expected by
Leonhard Euler, with a period of 435 days suggests a good
approximation of $\epsilon$ too.
This means that the suggestion of Eq.(\ref{epsilon}) is at least
workable for the equilibrium figure of the Earth\footnote{%
Among solid astrophysical objects, only the Earth is investigated
mostly extensively. One has then to learn from solid geophysicists
in order to develop a comprehensive astrophysical model for solid
quark stars.
}.%

However, strain energy develops when a solid quark star spins
down, which is proportional to $[\varepsilon_0(P_0)-\epsilon]^2$.
The first glitch of the quark star occurs when the strain energy
reaches a critical value \citep{z04}. Subsequent glitches take
place too as the stellar stress increases to critical points.
We note that the star's ellipticity decreases after glitches due
to (i) the release of strain energy and (ii) possible increase of
mean density if stellar volume shrinks as the star cools.

A star's ellipticity could be observed if the star's free
precession mode is discovered due to the relation $\epsilon\simeq
P/P_{\rm prece}$, where $P_{\rm prece}$ is the precession period.
Evidence for free precession for PSR B1828-11 was provided by
\cite{sls00}: periodic timing variation and changes in pulse
duration.
The derived ellipticity is $\sim P/P_{\rm prece}\simeq 0.405~{\rm
s}/500~{\rm d} \simeq 10^{-8}$ {\em if} the precession is really
in a free mode. But, based on Eq.(\ref{varepsilon0}), the
ellipticity is $\varepsilon_0(405~{\rm ms})\simeq 2\times
10^{-6}$, which is 2 orders larger than that expected.
This conclusion conflicts with Eq.(\ref{epsilon}). Why is PSR
B1828-11 so ``round''?
An answer proposed here is that this observed precession mode
could actually be torqued by external force, rather than free. In
fact, a disk-forced precession model for PSR B1828-11 was provided
by \cite{qxx03}, but other source torques by, e.g., magnetodipole
radiation or planet(s), are also possible. The precession period
could be very long (and thus results in a much small ellipticity
if assuming free precession) if torques are not strong. Due to the
torque exerted by the Moon and the Sun, the spinning Earth shows a
forced precession period of 25,700 yr, being much larger than 435
d, the period of free precession.
By the way, planet-torque induced precession modes could be
effective to search planets around pulsars.
%
%
%
%
Nevertheless, a free precession mode could have been discovered in
a bursting radio source, GCRT J1745-3009, in the direction of the
Galactic center region \citep{zx05}.
The ellipticity derived from $P/P_{\rm prece}$ is $\sim 10/77
\simeq 2\times 10^{-6}P_{\rm 10ms}$. Based on
Eqs.(\ref{varepsilon0}) and (\ref{epsilon}), the spin period
should be $P>0.1$ s, which is very reasonable for normal pulsars.
If the pulsar's ellipticity tracks almost $\varepsilon_0(P)$ due
to glitches, $\varepsilon>\sim \varepsilon_0(P)$, then its
ellipticity $\epsilon>\sim 2\times 10^{-5}$ and period $P>\sim
0.1$ s.
Statistically, a pulsar with $P>\sim 0.1$ s could be a radio
nulling (even extremely, with null fraction $>90\%$) pulsar
\citep{Biggs92}, which fits the model provided by \citep{zx05}.

A pulsar must be non-axisymmetric in order to radiate
gravitationally. A wobbling pulsar, either freely or forcedly, may
thus radiation gravitational waves.
This wave results in a perturbed metric $h_{\mu\nu}$ of space-time
($g_{\mu\nu}=\eta_{\mu\nu}+h_{\mu\nu}$), which is order of $h_0$
being given by \citep{h0,ja02},
\begin{equation}
h_0={128\pi^3G\rho_0\over 15c^4}\cdot {\epsilon R^5\over
dP^2}\theta \approx
2.8\times 10^{-20} \epsilon R_6^5d_{\rm kpc}^{-1}P_{\rm 10ms}^{-2}\theta,%
\label{h0}
\end{equation}
where approximations $I\simeq 0.4MR^2$ and $M\simeq 4\pi
R^3\rho_0/3$ are applied for solid quark stars in the right
equation, the pulsar's distance to earth is $d=d_{\rm kpc}\times
1$ kpc, $\theta$ is the wobble angle.
LIGO is sensitive to hight frequency waves, which recently puts
upper limits on $h_0$ for 28 known pulsars through the second LIGO
science run \citep{LIGO2005}. The upper limits are order of
$10^{-24}$, which means approximately an limit of $\epsilon
R_6^5<10^{-4}$. This observation may not conflict with
Eqs.(\ref{varepsilon0}) and (\ref{epsilon}) if the pulsars' radii
are slightly smaller than 10 km. In fact, possible candidates for
low-mass quark stars with small radii are proposed by
\cite{xu05a}.
Only three normal pulsars are listed in Table 1 of
\cite{LIGO2005}: the Crab (B0531+21), J1913+1011, and B1951+32,
others are millisecond pulsars.

It is conventionally suggested that millisecond pulsars are
recycled, with a spinup history during binary accretion, and the
ellipticity may not follow Eq.(\ref{epsilon}) since
$\varepsilon_0(P)>\epsilon$ for such stars during their spinup
phases.
Nonetheless, since millisecond pulsars are now in a spindown
phase, we may simply suggest $\varepsilon_0(P)\simeq \epsilon$,
which is valid for the Earth, and calculate the upper limits of
pulsars' radii and thus masses according to
Eqs.(\ref{varepsilon0}) and (\ref{h0}). Note that
Eq.(\ref{epsilon}) is still effective if millisecond pulsars are
born during an AIC (accretion-induced collapse of white dwarfs)
process \citep{xu05a}.
The upper limits of pulsars' radii and  masses are calculated,
which are presented in Table 1. Large radius limits ($>10$ km; and
thus high mass limit, $>M_\odot$) are for the three normal
pulsars, which does not show any constrain on the equation of
state since pulsar's maximum mass is only $\sim 2M_\odot$.
However, the LIGO's observation suggests a small limit of $R\cdot
\theta^{1/5}$ (a few kilometers), and thus a small limit of
$M\cdot \theta^{3/5}$ (between $10^{-2}M_\odot$ and
$10^{-3}M_\odot$), for those millisecond pulsars.
\begin{table}
\tabcolsep 3pt
\begin{tabular}{lrccrl}
~~~~pulsar                 & period~~   &       distance    &$h_0$& radius$\cdot \theta^{1/5}$ & ~~mass$\cdot \theta^{3/5}$\\
~~~~name       & $P$ (ms) &$d$ (kpc)&$(10^{-24})$& \,(km)~~~~ & $~~~~(M_\odot)$ \\
\hline
 B0531+21$^*$    &   33.08 &   2.00 &   $41 $ & 19.6~~~~~ & ~~~~6.4\\
 J1913+1011$^*$  &   35.91 &   4.48 &   $51 $ & 18.6~~~~~ & ~~~~5.5\\
 B1951+32$^*$    &    39.53 &   2.50 &   $48 $ & 22.4~~~~~ & ~~~~9.4\\
 B0021$-$72C &   5.76 &   4.80 &   $4.3$ & 2.6~~~~~ & ~~~~0.015   \\
 B0021$-$72D &   5.36 &   4.80 &   $4.1$ & 2.4~~~~~ & ~~~~0.012  \\
 B0021$-$72F &   2.62 &   4.80 &   $7.2$ & 1.5~~~~~ & ~~~~0.0030  \\
 B0021$-$72G &   4.04 &   4.80 &   $4.1$ & 1.9~~~~~ & ~~~~0.0061  \\
 B0021$-$72L &   4.35 &   4.80 &   $2.9$ & 1.9~~~~~ & ~~~~0.0059  \\
 B0021$-$72M &   3.68 &   4.80 &   $3.3$ & 1.7~~~~~ & ~~~~0.0043  \\
 B0021$-$72N &   3.05 &   4.80 &   $4.0$ & 1.5~~~~~ & ~~~~0.0031  \\
 J0030+0451      &   4.87 &   0.23 &   $3.8$ & 4.1~~~~~ & ~~~~0.056  \\
 J0711$-$6830    &   5.49 &   1.04 &   $2.4$ & 3.0~~~~~ & ~~~~0.023  \\
 J1024$-$0719   &   5.16  &   0.35 &   $3.9$ & 3.9~~~~~ & ~~~~0.051  \\
 B1516+02A       &  5.55 &   7.80 &   $3.6$ & 2.2~~~~~ & ~~~~0.0090   \\
 J1629$-$6902    &   6.00 &   1.36 &   $2.3$ & 3.0~~~~~ & ~~~~0.024  \\
 J1721$-$2457    &   3.50  &   1.56 &   $4.0$ & 2.1~~~~~ & ~~~~0.0084  \\
 J1730$-$2304&   8.12 &   0.51 &   $3.1$ & 5.0~~~~~ & ~~~~0.11  \\
 J1744$-$1134&   4.07  &   0.36 &   $5.9$ & 3.5~~~~~ & ~~~~0.037  \\
 J1748$-$2446C   &   8.44 &   8.70 &   $3.1$ & 2.9~~~~~ & ~~~~0.021  \\
 B1820$-$30A &   5.44 &   7.90 &   $4.2$ & 2.2~~~~~ & ~~~~0.0094   \\
 B1821$-$24  &   3.05 &   4.90 &   $5.6$ & 1.6~~~~~ & ~~~~0.0037   \\
 J1910$-$5959B   &   8.36 &   4.00 &   $2.4$ & 3.2~~~~~ & ~~~~0.028  \\
 J1910$-$5959C   &   5.28 &   4.00 &   $3.3$ & 2.4~~~~~ & ~~~~0.011  \\
 J1910$-$5959D   &   9.04 &   4.00 &   $1.7$ & 3.2~~~~~ & ~~~~0.028  \\
 J1910$-$5959E   &   4.57 &   4.00 &   $7.5$ & 2.5~~~~~ & ~~~~0.013  \\
 J1939+2134  &   1.56 &   3.60 &   $13 $ & 1.2~~~~~ & ~~~~0.0015  \\
 J2124$-$3358&   4.93 &   0.25 &   $3.1$ & 3.9~~~~~ & ~~~~0.048  \\
 J2322+2057  &   4.81 &   0.78 &   $4.1$ & 3.2~~~~~ & ~~~~0.027  \\
  \hline
\end{tabular}
\caption{The upper limits of the radii and masses of 28 pulsars
targeted in the second science run of LIGO, which depend on the
wobble angle $\theta$. Normal pulsars listed are starred (*),
others are millisecond pulsars. The $h_0$ limits are from
\cite{LIGO2005}, while the pulsar data are obtained through
http://www.atnf.csiro.au/research/pulsar/psrcat/.}
\end{table}
This hints either a low mass ($\ll M_\odot$) of pulsars or a small
wobble angle $\theta$.
Factually, in order to explain its polarization behavior of radio
pulses and the integrated profile (pulse widths of main-pulse and
inter-pulse, and the separation between them), the
fastest\footnote{%
A 1.39 ms radio pulsar could have been discovered in the globular
cluster Terzan 5 by the Robert C. Green Bank Telescope (GBT).
} %
rotating millisecond pulsar PSR J1939+2134 (B1937+21) is supposed
to have mass $<0.2M_\odot$ and radius $<1$ km \citep{xxw01}. This
conclusion does not conflict with the value in Table 1.
Observationally, the precession angle (i.e., the angle between the
spin vector and total angular momentum), which is smaller than
$\theta$, should be smaller than the pulsar radio beam angle (a
few tens of degrees).
The radius upper limits for the millisecond pulsars could be about
two times the values listed in Table 1, and then the mass limit of
($3\sim 10$) times, if $\theta\simeq (1^{\rm o}\sim 10^{\rm o})$.
Actually, a small radius of pulsar-like stars is not surprising
since we have detected six Central Compact Objects in supernova
remnants (CCOs) with black-body radius of ($0.3\sim 2.4$) km
\citep{cco}, and seven Dim Thermal Neutron stars (DTNs) with
possible radius of a few kilometers \citep{dtn}.

{\em Elliptic deformation v.s. bumpy distortion.}
The lack of symmetry about a pulsar's rotation axis could be the
result of either elliptic deformation because of  rotation, which
is discussed in \S2 and \S4. or bumpy distortion (i.e., localized
mountains on stellar surface).
Strong magnetic field is suggested for producing the bumps, since
the magnetic and rotating axes are generally not aligned. However
for quark stars, this mechanisms may not work due to a relatively
negligible magnetic force, $(B^2/8\pi)/(\rho_0c^2)\sim
10^{-5}B_{16}$, even for fields ($B=B_{16}\times 10^{16}$ G) as
strong as $10^{16}$ G.
Glitches of solid quark stars could produce bumps, with a maximum
ellipticity \citep{owen05},
\begin{equation}
\epsilon_{\rm max} \sim 10^{-3} ({\sigma_{\rm max}\over
10^{-2}})R_6^{-6}(1+0.084R_6^2)^{-1},
\label{emax}
\end{equation}
where $\sigma_{\rm max}$ is the stellar break strain. This
ellipticity is larger for low-mass quark stars due to weaker
gravity. But for normal neutron stars, \cite{owen05} derived the
maximum elastic deformation, $\epsilon_{\rm max}$, induced from
shear stresses is typically only $6.0\times 10^{-7}$. The real
ellipticity of a quark star could be far from $\epsilon_{\rm
max}$, but may be approximately $\varepsilon_0(P)$, due to stress
releases through star-quake induced glitches \citep{z04}.

\section{Conclusions}

Detection of gravitational waves can certainly test the general
theory of relativity, and also open a new window for us to observe
astrophysical phenomena.
Observations in gravitational wave band may reveal the nature of
pulsar-like stars, and help to answer numerous questions: Are
pulsar-like stars normal neutron stars or quark stars? Is quark
matter with hight density but low temperature in a solid state? Do
quark stars with low masses exist in the Universe?
The paper is summarized as follows.

(1). During the birth of quark stars, quark nuggets may be
ejected, and the global quadrupole oscillation around ellipticity
$\varepsilon_0(P)$ [see Eq.(\ref{varepsilon0})] could also be a
source for gravitational waves.

(2). No gravitational wave originated from $r-$mode instability
could be detected for pulsars to be quark stars with low masses or
in a solid state, for isolated as well as accreting compact stars,
even the compact stars spin at sub-millisecond periods.

(3). Through the second LIGO science run, we may find upper limits
of masses and radii for millisecond pulsars, although one can not
constrain on the masses and radii of normal neutron stars. The
radius of PSR B1937+21 could be smaller than $\sim 2$ km if its
wobble angle $\theta$ is between $1^{\rm o}$ and $10^{\rm o}$.
Future LIGO observations could put tight constraint on $R\cdot
\theta^{1/5}$ and thus $M\cdot \theta^{3/5}$ if pulsar-like stars
are actually solid quark stars.

(4). We suggest the precession mode discovered in PSR B1828-11 is
not free, but could be forced by fossil disk or planets. The
geometrical parameters of torqued precession might be obtained by
precise timing of radio pulses. The bursting radio source, GCRT
J1745-3009, could be a freely precessing radio pulsar with spin
period $\sim 0.1$ s, which could be tested by searching pulsed
emission from the source.

As the mass-radius relation for low-mass quark stars ($M\propto
R^3$) is in striking contrast to those of normal neutron stars
with low masses ($M\propto R^{-3}$), gravitational radiation from
low-mass neutron stars and quark stars should be very different
for processes of both fluid instability and stellar wobbling. We
have just studied mass-dependent waves for quark stars, but it is
also necessary to obtain the mass-dependent behavior for normal
neutron stars, in order to put theoretical models directly in
front of gravitational wave observations.
Additionally, similar to the case of double white dwarf binaries
\citep{svn05}, detecting gravitational wave via LISA (the Laser
Interferometer Space Antenna) from binary quark stars is also
valuable to constrain the masses and radius. The radiation
reaction, tides, and mass transfer are very different for binary
low-mass quark stars, white dwarfs, and normal neutron stars.


{\em Acknowledgments}:
I would like to appreciate various stimulating discussions in the
pulsar group of Peking university and to thank the anonymous
referee for his/her valuable suggestion.
This work is supported by National Nature Sciences Foundation of
China (10273001) and by the Key Grant Project of Chinese Ministry
of Education (305001).

\end{document}